\pdfoutput=1
\documentclass[a4paper,fleqn,usenatbib]{mnras}

\usepackage{latexsym,graphicx}
\usepackage{graphics}
\usepackage{epic}
\usepackage[varg]{txfonts}
\usepackage{mflogo}
\usepackage{epsfig}
\usepackage{eepic}
\usepackage{rotating}
\usepackage{lscape}
\title[Connection between CNSFRs and active nuclei in NGC 1068]{Connection  between the   Circumnuclear Star-Forming
 Regions and the active nuclei in NGC 1068}

\author[Villica\~{n}a-Pedraza, I. et al.]{
Villica\~{n}a-Pedraza I.$^{1}$,\thanks{E-mail: astrojupiter62@hotmail.com}
Mast D.$^{2}$,
D\'iaz A.$^{1}$,
Carreto-Parra F.$^{3}$,
Hagele G.$^{4}$,
Cardaci, M.$^{5}$
\\
$^{1}$Departamento de F\'isica Teorica, Universidad Aut\'onoma de Madrid, Spain. \\
$^{2}$Observatorio Astron\'omico de Cordoba, Argentina.\\
$^{3}$Las Cruces, New M\'exico State University, USA.\\
$^{4}$ CONICET at UNLP-Universidad Nacional de La Plata.\\
$^{5}$ Facultad de Ciencias Astronómicas y Geofísicas, CONICET.}

\pubyear{2016}

\begin{document}
\label{firstpage}
\pagerange{\pageref{firstpage}--\pageref{lastpage}}
\maketitle

\begin{abstract}
{In recent years, Circumnuclear Star Forming Regions (CNSFRs) have been studied in different frequencies and telescopes, diverse studies about
connection between nuclear activity and star formation in the CNSFRs were doing using optical data, IFS and models.} 
{We obtain data of the Seyfert 2 galaxy NGC 1068 at J band between  1.35-1.32$\mu$ using the 8.1m Gemini telescope and GNIRS IFU mode.}
{We found the radial velocity and intensity maps for  [FeII]$\lambda$12570, Pa$\beta$ $\lambda$12814 and FeII $\lambda$13201 lines of all the components.}
{The  gas emission has a double origin  associated with photoionization  by young stars or excitation by collisions, we found a high influence of shocks produced by the jet.
The intensity of Pa$\beta$ emission show that this component is ionized by a younger cluster or more massive stars. From the analysis of the velocity fields,
the data related with the component one of Pa$\beta$ are consistent with the presence of rotation which is displaced from the maximum continuous, our observations may  point to
the presence of a bar.}

\end{abstract}


%%%%%%%%%%%%%%%%%%%%%%%%%%%%%%%%%%%%%%%%%%%%%%%%%%

%%%%%%%%%%%%%%%%% BODY OF PAPER %%%%%%%%%%%%%%%%%%

\section{Introduction}

The  Circumnuclear Star-Forming
 Regions (CNSFRs) with sizes until hundred of parsecs (\citet{Diaz2000}) are constituted by several HII regions ionized by compact and bright
stellar clouds. Using stellar velocity dispersions (39-67 km/s), the previous authors derived the dynamic masses for the complete stellar formation complex
and the individual stellar clouds with masses from 4.9x10$^{6}$Mo to 4.3x10$^{7}$Mo.  

There is evidence of the presence of at least two different ionized kinematic gas components in the regions. The narrow component provided by setting two 
Gaussian components seems to have a relatively constant value for all CNSFRs studied, with an average of 23 km/s. This narrow component could be identified with ionized gas on a
rotating disc, while the stars and the gas fraction responsible for the wide component seem more related to the regions of star formation.

Furthermore, \citet{Diaz2000} have investigated the possible connection between nuclear activity and star formation in circumnuclear galaxies CNSFRs  with 
variable degrees of nuclear activity, concluding that there is no relationship in the formation of stars  and the region of nuclear activity caused by remnants
of supernovae with gas and radiation expelled from the active nucleus. \citet{Dors2008}  combined optical IFS data (Integral Field Unit) 
Gemini and models to determine the abundance of gas and star forming rate on the CNSFRs of two active galaxies NGC1097 and NGC6951. 

 \citet{Riffel2008} and \citet{Riffel2010}  using slicers inside GEMINI near infrared ( NIFS and GNIRS in the IFU mode ) studied the AGN - starburst connection and
Kinematics of gas in galaxies NGC4051 active and NGC7582 , respectively. They concluded that there are two forms of connection: first, molecular gas leads to the formation 
of stars in the circumnuclear region; second, the molecular gas is formed in the  circumnuclear star formation region.

The galaxy NGC 1068 has been long considered as the prototype Seyfert 2 galaxy, but it shows wide components lines (FWHM of the order of 3500 km/s), typical emission lines
of Seyfert 1 when its spectrum is obtained in polarized light  \citet{Antonucci1985}, \citet{Antonucci1993} but part of the unpolarized light reaching 
the observer has been absorbed by the toroid. It has also been classified morphologically as a spiral galaxy of type Sab.It has angular size of 7.1'x 6'. It is located at a
distance of 14.4Mpc  \citet{Ulvestad1987} and has a redshift  z=0.00379 ± 0.00001  \citet{Huchra1999} .The position of the active
nuclei in far Infrared is R.A. 02 42 40.771, DEC(J2000) -00 00 47.84  \citet{Skrutskie2006}.

The mass of the black hole, derived from the observational data is about 8x10$^{6}$Mo  \citet{LodatoBertin2003}. Observations in wavelengths the optical-UV 
show that the emission is concentrated in a cone that extends to the northeast  \citet{Capetti1995}, \citet{Colbert2002}. 
The same result was detected in maps [OIII]  \citet{Ruiz2001} and [NII]  \citet{Cecil1990}. The direction of polarized light 
cone coincides with jets observed in radio maps obtained with VLA  \citet{WilsonUlvestad1982}. The Cone also is seen in the southwest direction in 
optical-ultraviolet and [OII] but weaker. A stellar bar of 2Kpc was detected in the near infrared  \citet{Scoville1988}, in turn,
surrounded by a starburst circumnuclear it has also been detected in molecular gas lines  \citet{Sternberg1994} and continuous radio 
 \citet{Gallimore1996}. The emission of this burst added to other star formation regions contribute the order half of the bolometric 
 luminosity  of the galaxy  \citet{Davies1998},  \citet{JimenezBailon2004}.
Optical spectroscopy  made with the HST (Hubble Space Telescope) showed that NGC 1068 has a region of complex coronal lines extended to hundreds of parsecs. 
``The dynamics of the  molecular gas in the torus show strong non-circular motions and enhanced turbulence superposed on a surprisingly slow rotation of the disk, 
evidence suggesting that the molecular torus is less inclined''  \citet{GarciaBurillo2016}.

NGC 1068 represents an ideal object to try to establish the origin of the different   kinematics components  of the   gas in circumnuclear regions.
The aim of this work is to study the field of the velocities of the circumnuclear stellar formation regions and its possible connection with the active kernel of the galaxy.
We will map field velocities of the gas in these regions with good spatial sampling, and we will separate the field velocity complexes of circumnuclear star formation,
to study radial velocity of different zones to understand their effect on a possible widening of emission lines on the gas and distinguish the origin of the two components.

\section{Observations and Reduction}

The spectroscopic data of NGC 1068 in infrared were obtained from the database of the 8.1m telescope  Gemini South with GNIRS instrument in its configuration integral
field spectrograph on the night of December 26, 2004. The seeing that night was between 0.6 and 0.8 arcsec in the band V. The instrument is set up in a position
angle PA = 0, ie,the slices are east-west orientation.

The data analyzed in this paper were obtained with the spectrograph (GNIRS). When the Information of this work was acquired spectrograph was on the 8.1m 
telescope Gemini South. GNIRS provides integral field spectroscopy with a spectral resolution from R = ( $\lambda$ / $\lambda$ $\Delta$x ) 4800 to 1600, J -band in a
visual image field of 4.8 x 3.2 arcsec$^{2}$.
The central wavelengths of the gratings  GNIRS can be adjusted reaching lengths of higher or lower wavelengths. For this a network of 32 l/mm is used that depending
on the angle; it can be adjusted from the X (1.10 $\mu$m) to the M band (4.80  $\mu$m), although generally is used in bands J , H , K (1.25  $\mu$m to 2.20  $\mu$m). 
The IFU- GNIRS divides the plane of the sky into 21 slices of 0.15 arcsecs wide each. 
Six exposures of 300 second were conducted in the observed zone of NGC 1068. Spectra flat screen field ( GCALflat ) and flat field of sky were also obtained
(Twilight) at exposures 2.5 sec. and 30 sec. respectively. It has also been obtained a spectrum of a calibration lamp ( Arc- Lamp) with a 1.5 sec exposure
to calibrate wavelength.

The reduction of these data was performed using the tasks of pack GNIRS of IRAF called GIRAF . In the process of reduction, the images were prepared to cut them 
(only if necessary with nscut), flat fields were created for screen and sky (nsflat) , the telluric star and object of science were normalized , correction of
distortion of light (nssdist) , identification of the lines of the lamp was made, the sky was subtracted , the spectra obtained in the same spot were combined,
the calibration was performed in wavelength  (nstransform). Finally, the 2D spectrum is extracted in a 3D cube data (nfcube). 
The cube contains spatial information in two directions and information Spectral on in the third. The final cube contains 651 spectra of the band J.

\section{MEASUREMENTS AND RESULTS}

\subsection{MEASUREMENTS}

The data cube contains 651 spectra obtained on J band. Each spectrum obtained has been manually analyzed component by component to keep 
valuable information that could help us understand in more detail the kinematics of NGC 1068. The spectrograph has 21 slices or mirrors. We obtained 31
spectra of each in order to complete the 651 mentioned. 
The spectra of NGC 1068 show emission lines of hydrogen recombination  Pa$\beta$ ($\lambda$12814) as well as [FeII] $\lambda$12570 and FeII ($\lambda$13201 ).
Like the optical spectrum \citet{Arribas1996} the emission lines in the infrared band show significant variation in their profiles with asymmetries and peaks of at least 
three components in many of them.

The maps were drawn from our measurement tables of the lines, creating a script that makes an order that allows  IRAF to convert the information into an image that will be
eventually a map. The maps are in coordinates x, y, z,  where z is the value of our variable for example flux , for x, y subtends 0.15 arc seconds in the sky per pixel.
The bar that accompanies each map indicates the value of each pixel by color.

Some of the spectra obtained in this work are shown (see Fig.1), note that the components of the lines are well defined, profiles so characteristic  of NGC 1068.
The obtained flow is in counts since they are not calibrated flow, i.e., we  don't have a standard to calibrate star flow, then the y axes in our spectra is the intensity 
measured by not being a calibrated flow. The equivalent width (Eqw) it is the intensity of the line relative to the continuum.

In general, as shown and mentioned above, the strongest lines are Pa$\beta$ ($\lambda$12814 ) and [FeII]$\lambda$12570 while FeII($\lambda$13201 ) , although it never 
disappears spectra , it becomes weak in most of the spectra. It is noteworthy that the Pa$\beta$ line in several spectra does not appear so could not be measured in all of them.
For to analyze each line with its components has been used IRAF task splot program adjusting at least two discrete components ( " d " for doublet , "g " for Gaussian fit).

\subsection{RESULTS}

In this section the results of this work show : Radial velocity of material in each region studied, flow lines and components; Intensity maps (flow)
for two components of each line ([FeII] $\lambda$12570 Pa$\beta$ $\lambda$ 12814 ); FWHM maps and Eqw ; Radial velocity maps; Location of the galactic nucleus on maps
with coordinates of Right Ascension (RA) and  Declination (DEC); understanding on behavior of the material in the study area.

The FWHM was disaffected of the instrumental width. To determine this we used the average of FWHM of the 10 most intense emission 
lines from the sky, obtaining a value 182 km/s . The instrumental width was quadratically substacted to the  FWHM values measured.

%\subsubsection{Results obtained for the emission line [ FeII ] λ12570, Paβ λ12814 and FeIIλ13201.}
The line  [FeII]$\lambda$13201 shows weak in most of the spectra, so that the mapping velocity and FWHM of that line would show a great error and it is not very reliable.

\subsubsection{[FeII]/Pa$\beta$ Ratios .}

In the map of figure 4 can be seing the ratio of lines [FeII]/Pa$\beta$. The ratio was made dividing the line intensities [FeII] and of Pa$\beta$ for its main component. 
The ratio allows separate regions where photoionization dominates. The most common values range are from 0.5 to 2. In this map (Fig. 4), the nuclear region 
(Marked with x) has values around 2. This would imply that there dominate the shocks, the SN, and the winds; but moving away from the region we can not say much
for poor signal/noise. 
The map for the second component is not shown as values indicate  little signal to noise.

\subsubsection{Spectra settings on maps}

Once done the maps of various line parameters of [FeII] and Pa$\beta$ it has been found the spectra configuration with respect to the maps. 
The maps are mapped  in 21x31 arrangements (y-axis vs x axis), the scale in arcsec is 0.15x0.15 by spaxel,square. Each box on the map corresponds to a spectrum.

\subsubsection{Error analysis}

The error map gives us an idea of the regions that are acceptably measures. For the component one of the map of Pa$\beta$, the purple regions are regions with errors smaller
than 50 km/s. Blue regions, they range from 50 to 100-150 (acceptable limit). To know what is the acceptable limit according to our data, we calculated the velocity resolution
and we have: if our spectrum for the line Pa$\beta$  at $\lambda$ = 12800 (approximately) has a Resolution on 3.85($\Delta$ $\lambda$) $\AA{}$, and " c" the speed of light,
we use the equation $\Delta$ $\lambda$ / $\lambda$ =$\Delta$V / c = 90km/s (each pixel in the spectrum). In figure 5 are error maps.

 %\subsection{Figures and tables}


\begin{figure}
	\includegraphics[width=5cm]{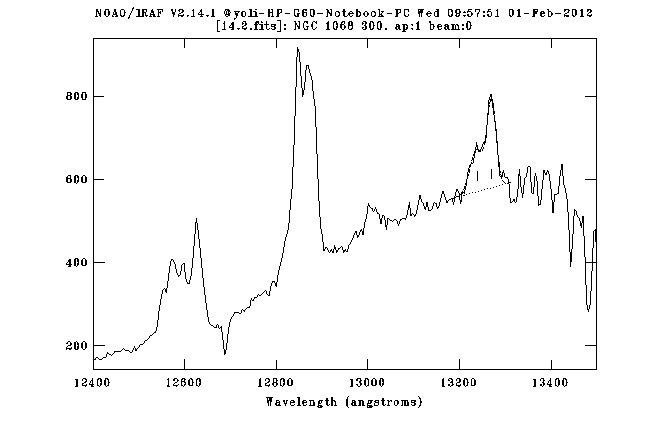}
	\includegraphics[width=5cm]{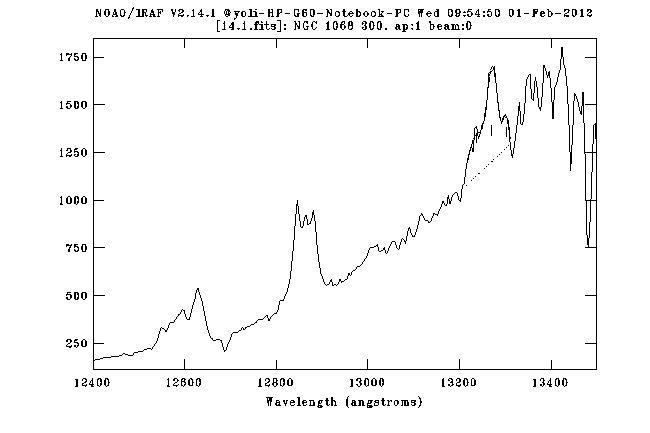}
	\includegraphics[width=5cm]{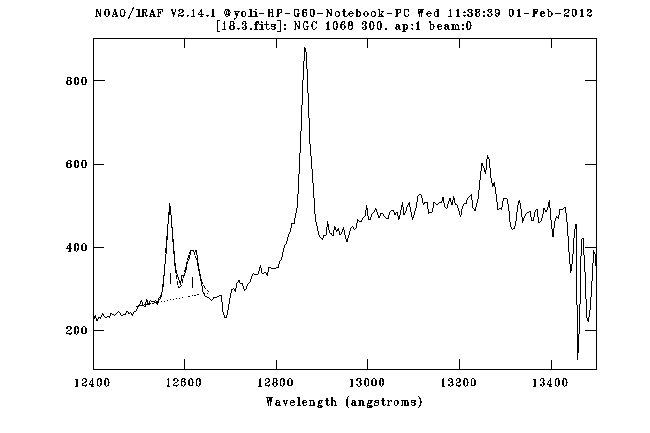}
	\includegraphics[width=5cm]{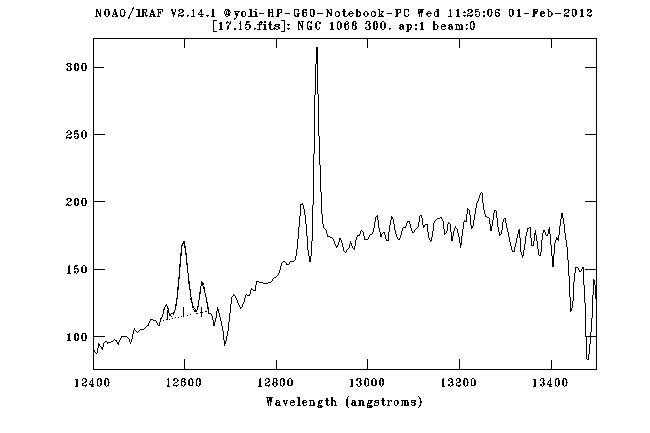}
	\includegraphics[width=5cm]{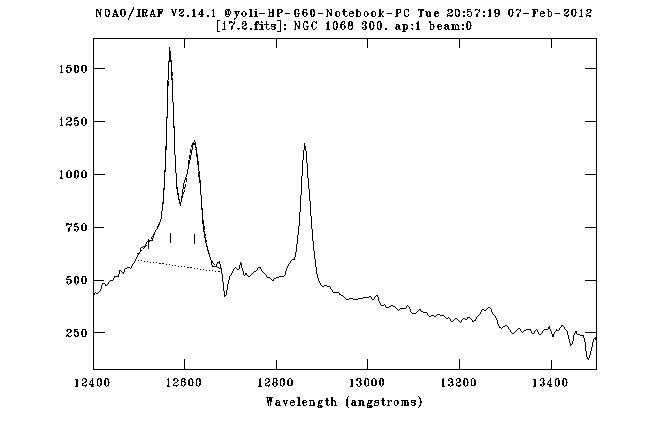}
	\includegraphics[width=5cm]{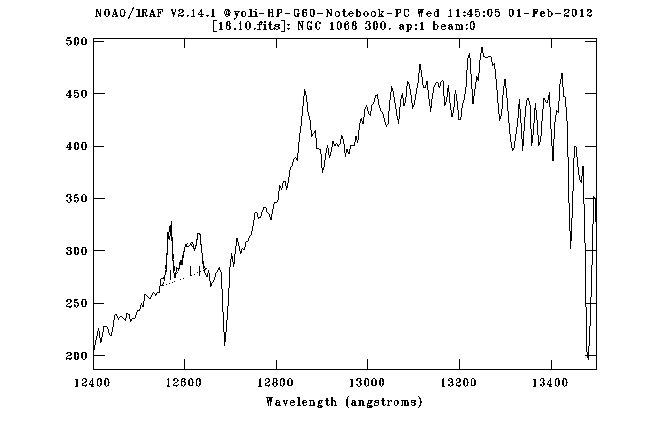}
	 \caption{Most representative spectra obtained 651 spectra which can be seen clearly
       three lines l of emission on that have been analyzed. El spectral range for the X axis ($\lambda$) is 12000-13000
       $\AA{}$. The Y axis is in accounts (intensity)}
    \label{fig:spectra}
\end{figure}

\begin{figure}
\begin{center}
\epsfig{file=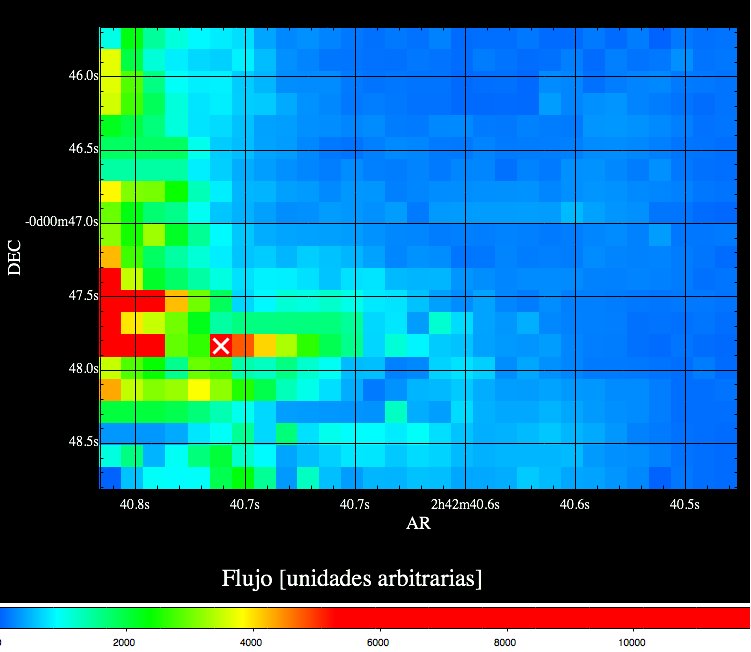,width=4cm}
\epsfig{file=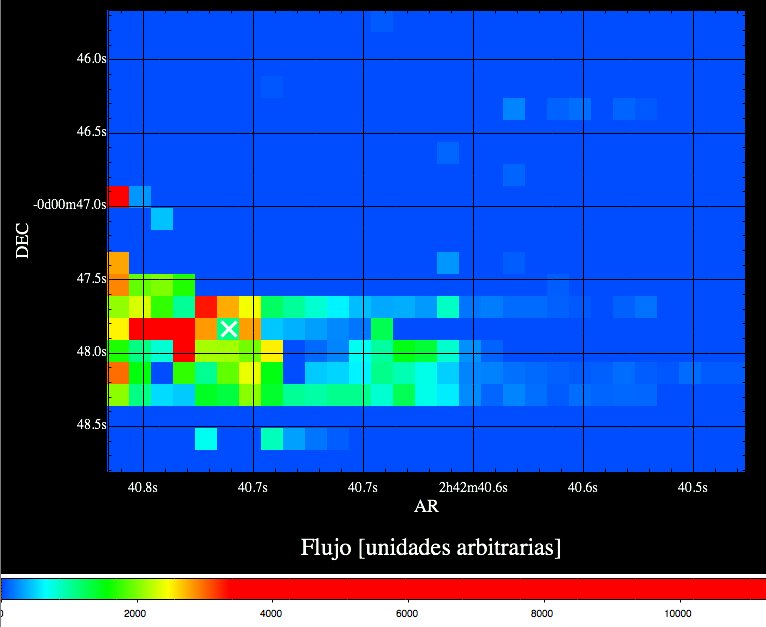,width=4cm}
\epsfig{file=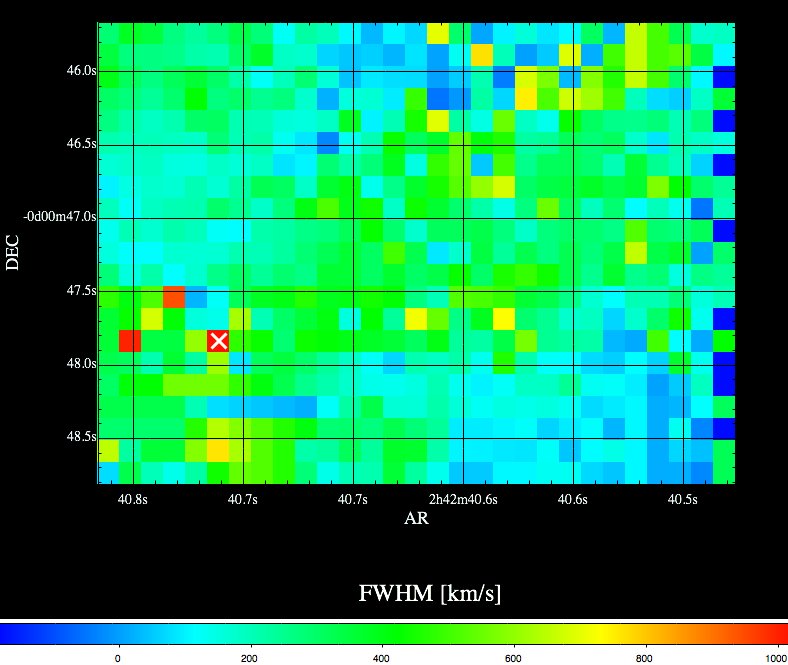,width=4cm}
\epsfig{file=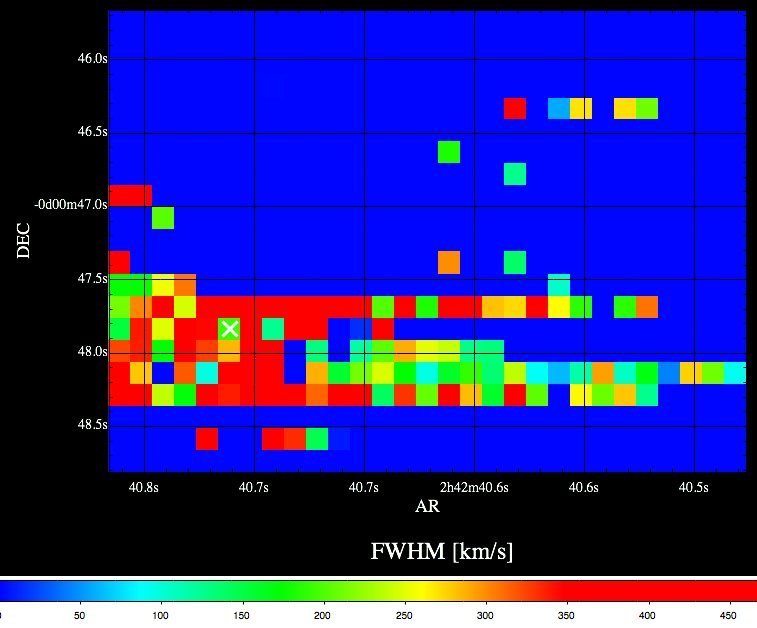,width=4cm}
\epsfig{file=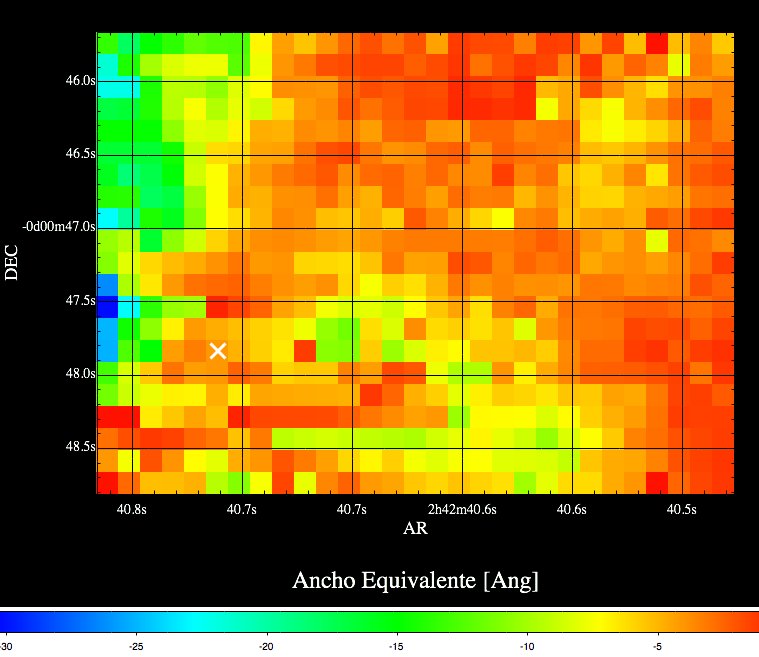,width=4cm}
\epsfig{file=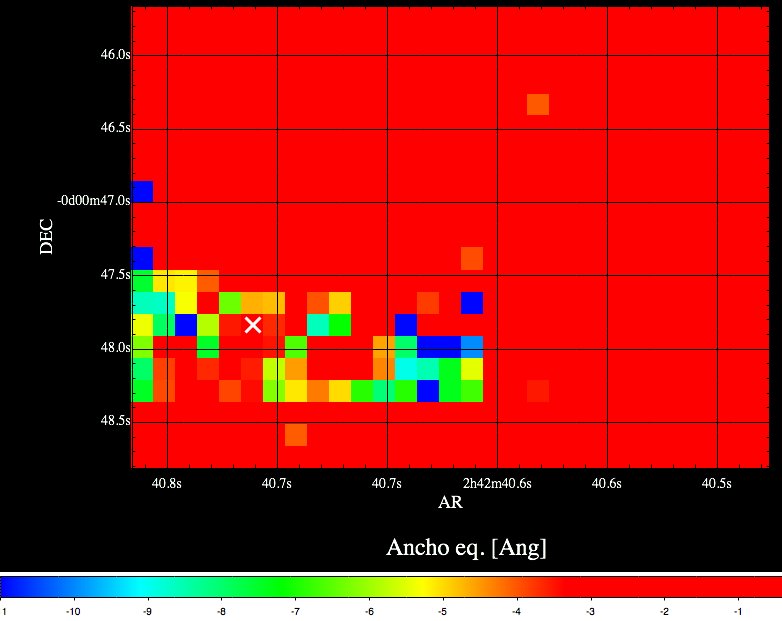,width=4cm}
\epsfig{file=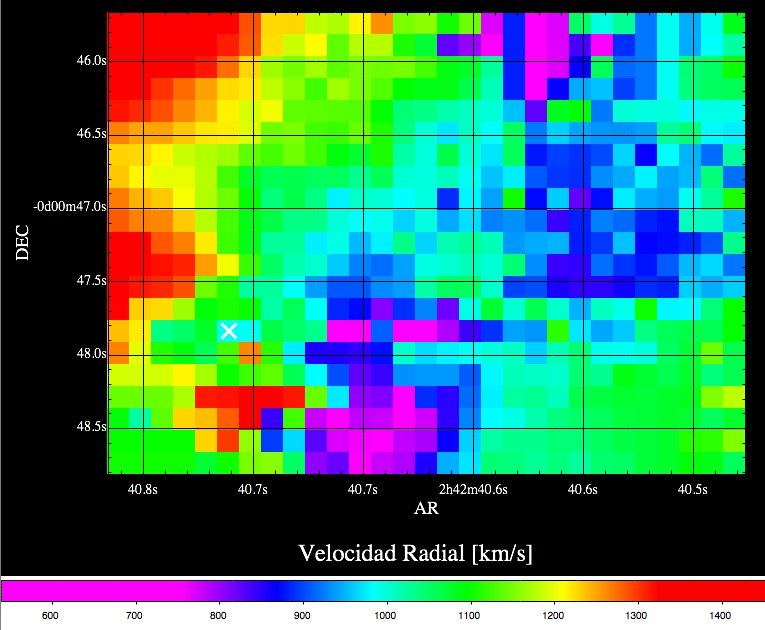,width=4cm}
\epsfig{file=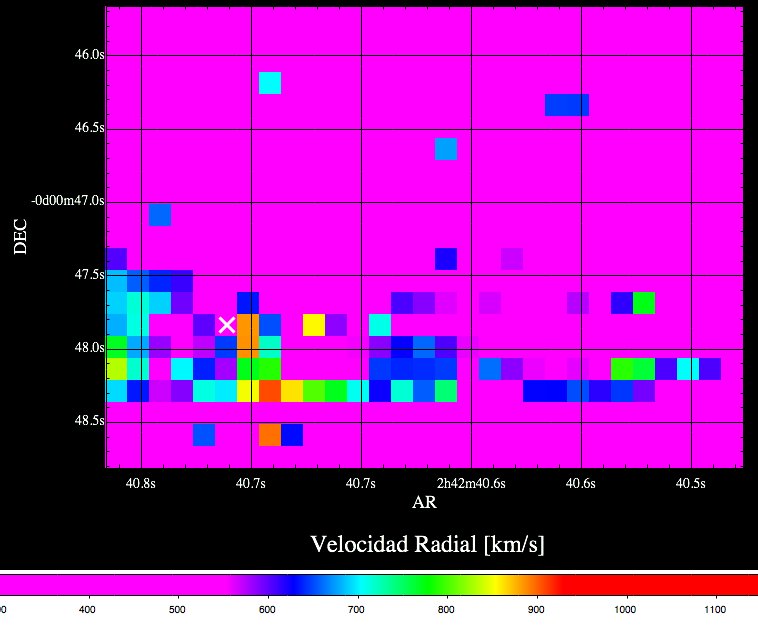,width=4cm}
\caption[Flux Maps,FWHM,EqW y  radial velocity for FeII]{\label{flux_FeII_comp1} {\footnotesize Flux maps for  FeII line
 $\lambda$12570 (one a two images); first block at center: FWHM maps in km/s; Centering second block, Eqw $\AA{}$; Finally,
 radial velocity maps in km/s. Color bar indicate parameter value. Left maps are for the first component, while, right maps to the second.
 Maps orientation is North up, east left. There are not maps for the third component because low signal to noise and was detected in a reduced region of the field.}}
\end{center}
\end{figure}

%\textbf{Mapas obtenidos para la l\'{\i}nea Pa$\beta$}

\begin{figure}
\begin{center}
\epsfig{file=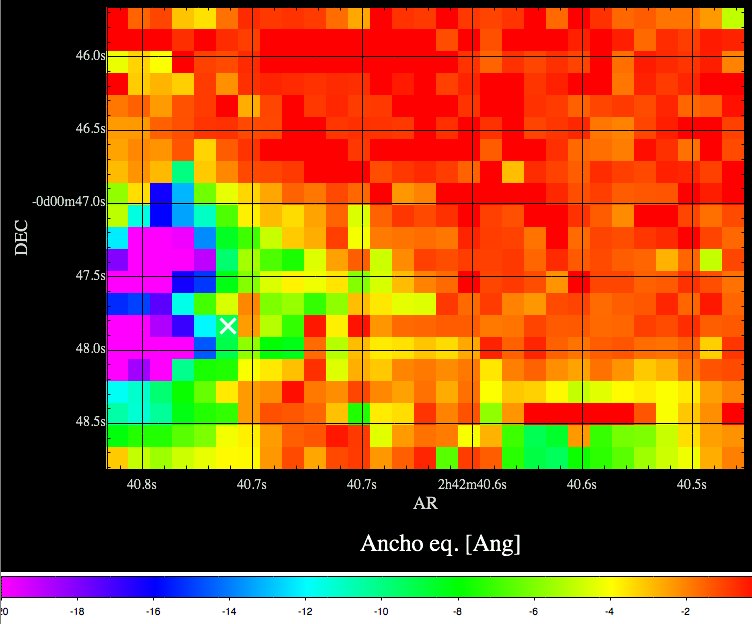,width=4cm}
\epsfig{file=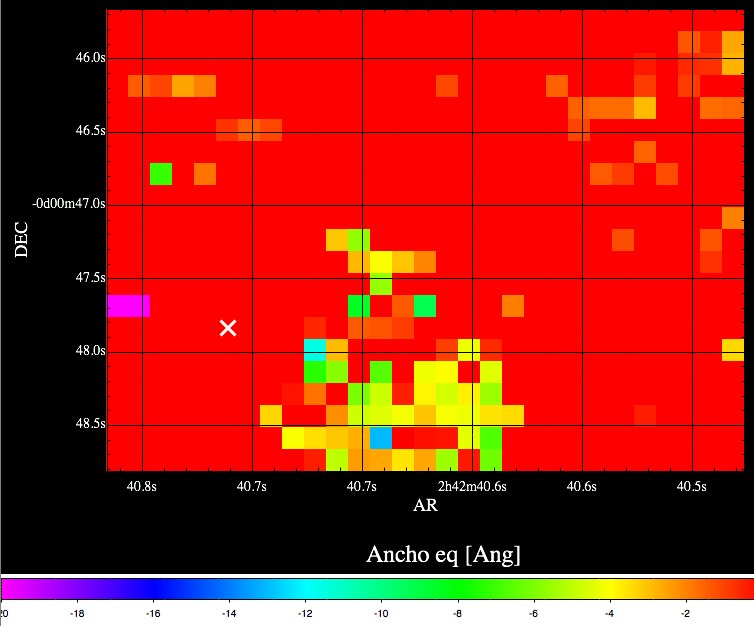,width=4cm}
\epsfig{file=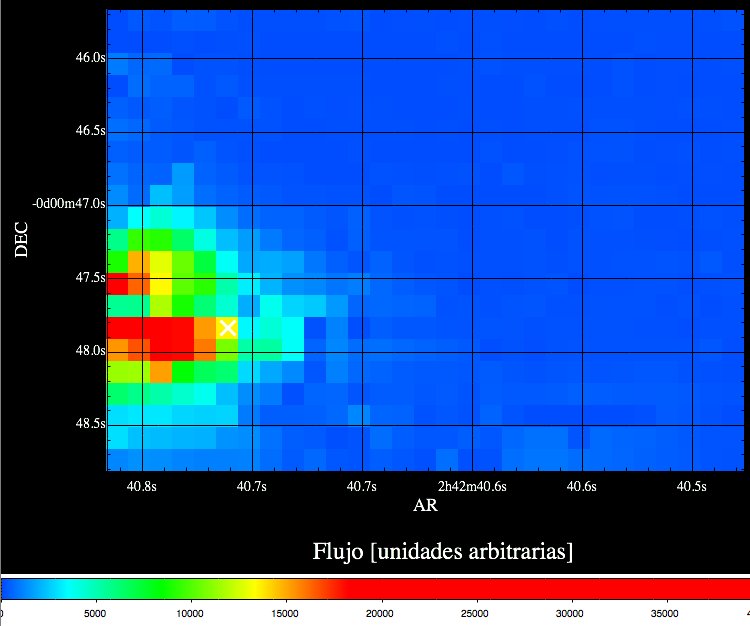,width=4cm}
\epsfig{file=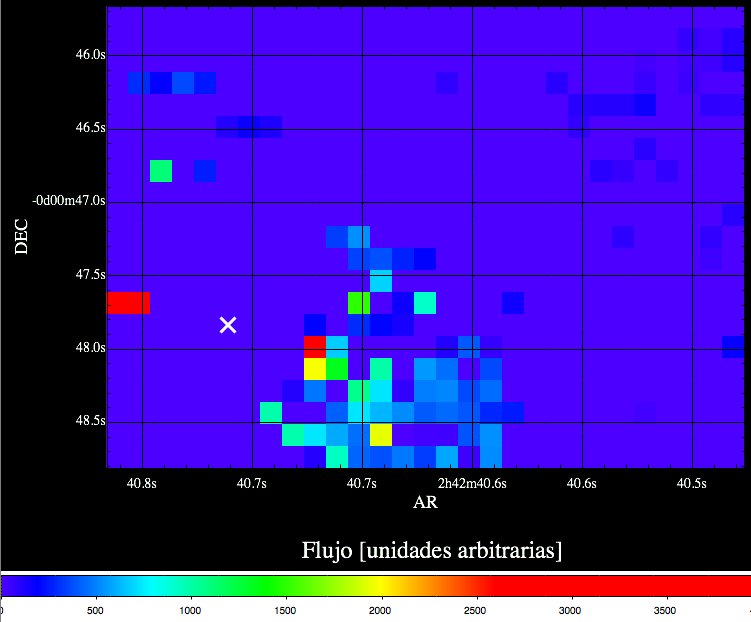,width=4cm}
\epsfig{file=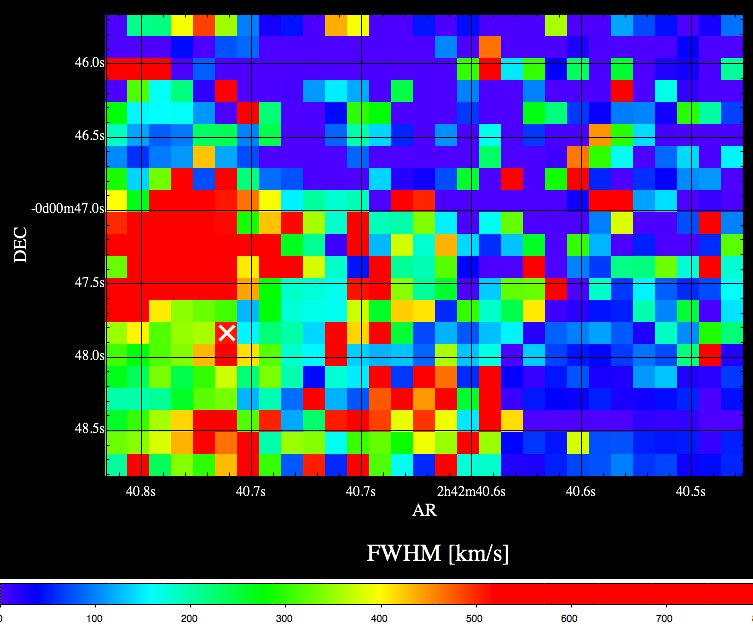,width=4cm}
\epsfig{file=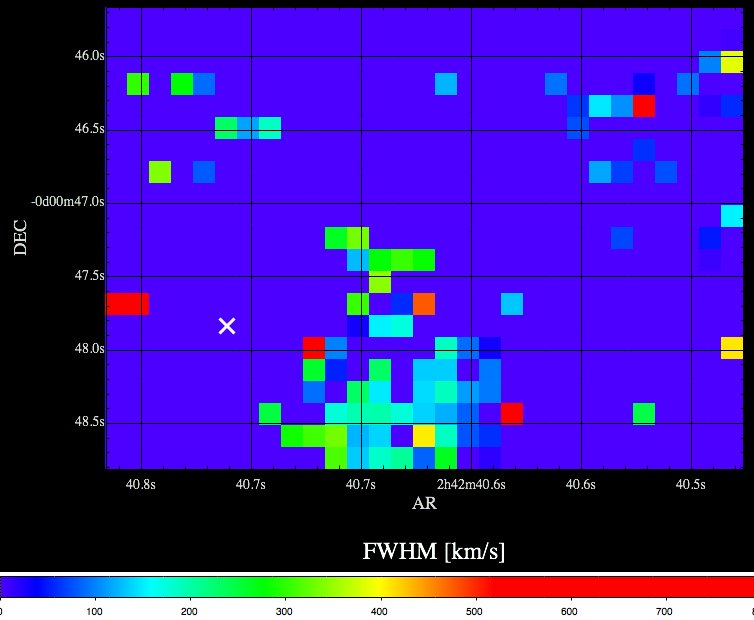,width=4cm}
\epsfig{file=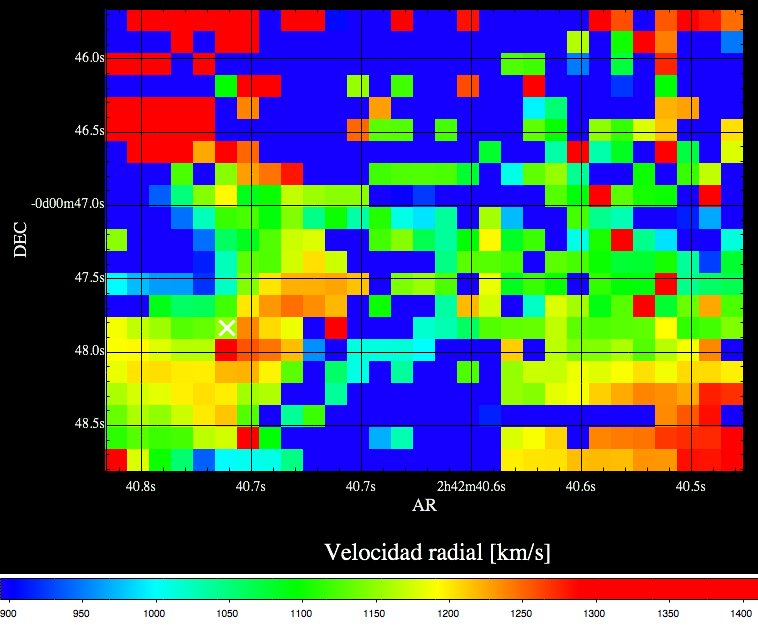,width=4cm}
\epsfig{file=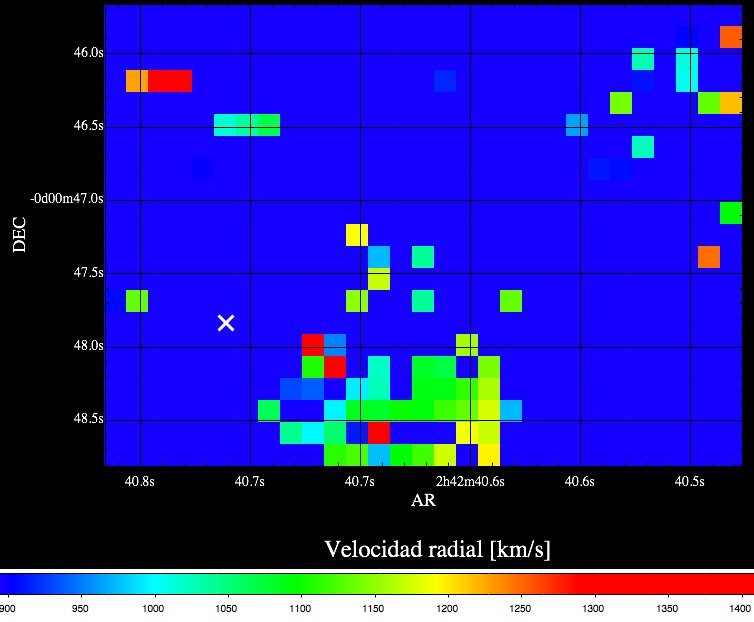,width=4cm}
\caption[EqW maps,flux,FWHM and Radial velocity of  Pa$\beta$ $\lambda$12814]{\label{EqW_comp1} {\footnotesize  Up: EqW  maps of the Pa$\beta$12814 line
in $\AA{}$; First block at the center, flux maps; Second block at the center, FWHM maps in km/s. Last one, Radial velocity maps in km/s.
Color bar indicate parameter value. Left maps are for the first component, while, right maps to the second.
 Maps orientation is North up, east left. There are not maps for the third component because low signal to noise}}
\end{center}
\end{figure}

\section{DISCUSSION}

\subsection{Intensities obtained in the Maps}
\normalsize
The intensity maps that have been obtained in this work for lines [FeII]$\lambda$12570 and $\lambda$12814 Pa$\beta$ are for main components. \citet{Riffel2011} have worked
with the Mrk 1157 galaxy observing with NIFS of Gemini telescope, making intensity maps for the same lines ([FeII] and Pa$\beta$ and its ratios, concluding that the emission 
flow for [FeII] increases due to shocks produced by the jet (radio), which destroys the dust grains expelling from the AGN. In addition \citet{Riffel2011}
worked with NGC 7582, which keeps recent star formation with a dark region in the central region \citet{Morris1985};. \citet{Regan1999};
\citet{StorchiBergmann2001}; \citet{SosaBrito2001}; \citet{Wold2006}, dust is responsible for asymmetrical contours flow maps obtained while the
star formation ring is circle. These clouds of star formation had not been observed in optical-IR. Comparing with the preceding information and maps obtained in 
this study we conclude that: i) emissions lines are elongated in the direction of the jet; ii) the emission of flow for [FeII] in both components is greater than for
Pa$\beta$ components, which means influence of shocks produced by the jet; iii) More intensity in the emission of Pa$\beta$ for component 1 would imply a
ionization region of the component 1 due to a younger cluster, or where there are more massive stars that ionize the gas.

\subsection{FWHM from maps}

\citet{Riffel2010} studied  NGC 7582, reporting the presence of a broad component of Br$\gamma$ line with FWHM of 4000 km/s. 
Another broad component observed (which disappeared) is h$\alpha$ and had also been reported by \citet{Aretxaga1999}, the latter seemed to have been caused by a transient
event and proposed three possible scenarios for its origin: i) Capture and breaking of a star by a supermassive black hole.
ii) A change in the redness of the toroid surrounding the core. iii) An explosion of Supernova Type II.

In this paper the map of FWHM for [FeII] has values ranging from 200-800 km/s being the most dominant about 600 km/s, indicating increased gas turbulence in
the area where that line is produced causing wider and deformed lines by turbulence. 

From the maps obtained in this work, we found that for  Pa$\beta$ line the FWHM  obtained for component one is higher than for the component two with values ranging from 
100-500 km/s and 100-300 km/s respectively with maximums at 300 km/s and 100 km/s respectively. We conclude that  exists more gas turbulence in 
this line for component 1 than in component 2.

\subsection{EqW from Maps}

In his study of NGC 7582 \citet{Riffel2010} found for Br$\gamma$ line an Eqw  of 30$\AA{}$ and  stellar formation of 5Myr.  \citet{Dors2008}  found similar values of CNSFRs
in NGC6951 and NGC1097 . A greater Eqw indicate more star formation as shown \citet{RodriguezZaurin2010}. For this paper, we can see that, in the case of the Eqw of 
[FeII]$\lambda$12570, we have values from 3-30 $\AA{}$ $^{-1}$. In the component one centered at 3$\AA{}$, and for the component two, values are similar. While in the case of
Eqw of Pa$\beta$  for component one, it has values from 1-6$\AA{}$ and for the component two, it has values dominated for 1-2$\AA{}$ .

Because the component 1 shows an EqW of  Pa$\beta$  bigger, we can estimate that this component would occur in an area with more 
 recent star formation or abundant.

\subsection{Ratio lines}

The map obtained can observe the ratio of the lines [FeII]/Pa$\beta$ . The ratio allows separating regions where photoionization and shock regions dominate.
The most common values are 0.5 to 2. In this map the nuclear region (marked with x) has values around 2. This implies that there dominate the shocks, the SN, 
and the winds; but moving away from the region we can not say much because poor signal/noise. The map for the second component is not shown because the values
indicate little signal to noise and higher errors.

\subsection{Radial velocity and dispersion velocity}

The optical spectra of the nuclear region show clear signs of middle-aged stars (Cid-Fernandes (2002)). 
About the kinematics of gas, the blue shift could be due to an inflow if the gas is located in the plane or a outflow if the gas is 
extended at high latitudes above the plane (conical structure facing us). \citet{Riffel2010} studied NGC 7582, the radial velocity field shows a blue shift
toward SE and a redshift NW, suggesting rotation although the kinematic center is shifted from the continuous  emission peak.
They concluded that the stellar velocity field shows distorted rotation which is offset relative to the core 50 pc (continuous peak); The distortion may be due to a bar 
in the nucleus. In this work the velocity for the two components are of 1200 km/ and 600 km/s for the line [FeII], while for the line Pa$\beta$ we have 1150 km/s and 700 km/s. 
In the work of \citet{Mazzalay2010} NGC 1068 was studied (with NIFS at 1.15-1.3 microns) but it was not adjust each of the lines of interest manually, it built codes to analyze them 
losing valuable  kinematics information from different components. \citet{Riffel2010} studied NGC 1066 with NIFS  (1.10-1.35 microns) to obtain a very similar
to that presented in this paper (GNIRS), but adjustments the same lines without considering the components, losing kinematics information of the galaxy.

\begin{figure}
\begin{center}
\epsfig{file=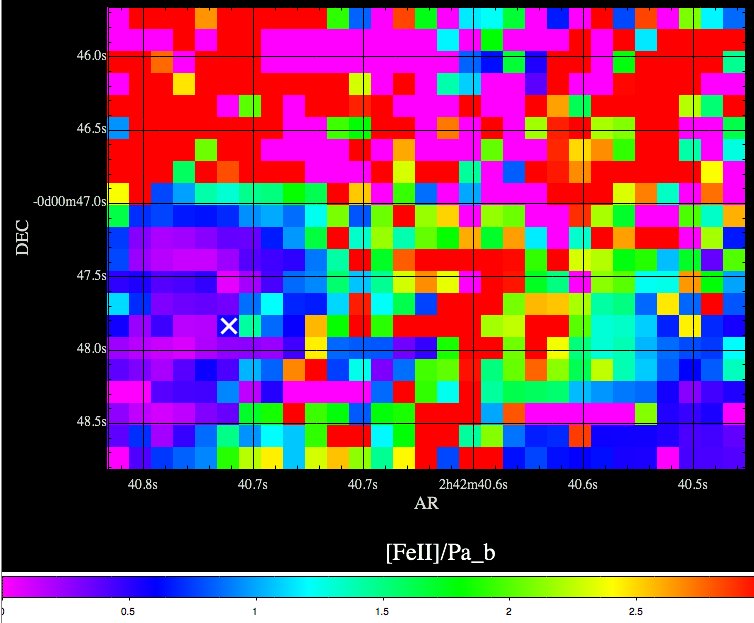,width=6cm} 
\caption[Ratio map of Intensity of FeII / Pa$\beta$]{\label{Cociente_Fe_Pa_comp1} {\footnotesize  Ratio map of Intensity of FeII/Pa$\beta$
using the principal component  of each line respectively. Orientation of the map is North up and east left.}}
\end{center}
\end{figure}

%\subsubsection{Spectra configuration}

%\begin{figure}
%\begin{center}
%\epsfig{file=mapa_arreglo.jpg,width=8cm,height=8cm} 
%\caption[Configuraci\'on de los espectros]{\label{mapa_arreglo} {\footnotesize Configuraci\'on de los espectros
%respecto de los mapas. Los mapas se encuentran cartografiados en arreglos de 21x31. El primer cuadro inferior izquierdo corresponde al espectro 1.1; 
%el cuadro superior izquierdo corresponde al espectro 21.1; El \'ultimo cuadro inferior derecho corresponde al espectro 1.31; el ultimo cuadro superior 
%derecho corresponde al espectro 21.31. A manera de ejemplo de los espectros que corresponden en algunas regiones se observa en la parte superior los espectros
%21.2, 21.15 y 21.29; en la parte media se muestran los espectros 12.2, 12.15 y 12.29; en la region inferior se observan los espectros 1.2, 1.15 y 1.29.}}
%\end{center}
%\end{figure} 

\begin{figure}
\begin{center}
\epsfig{file=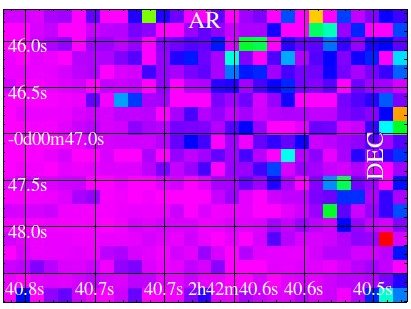,width=5cm} 
\epsfig{file=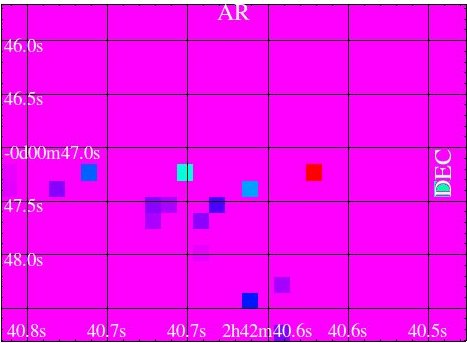,width=5cm} 
\epsfig{file=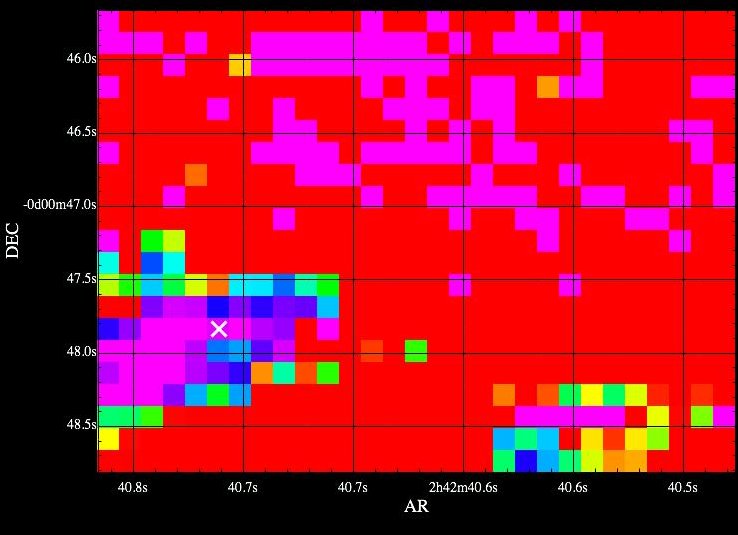,width=5cm} 
\epsfig{file=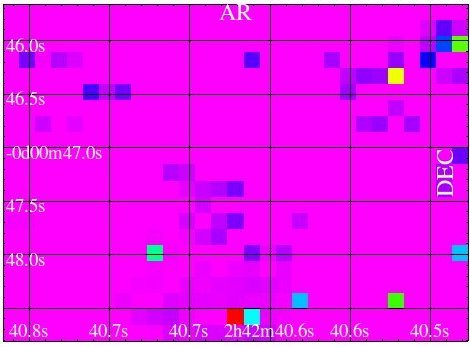,width=5cm}
\epsfig{file=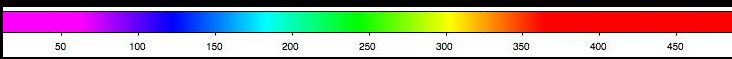,width=4cm}
\caption[Error analysis]{\label{Evel_FeIIc1} {\footnotesize Error maps  for radial velocity. On the Superior part are the error maps for [FeII].
Inferior part, the corresponding maps to the Pa$\beta$ line. The maps are show for two components, the first at left, and second at right.
The color bar represents the error value  km/s. To get error maps we use the Keel formula (1996).}}
\end{center}
\end{figure}  

\section{Conclusions}

In this paper, 651 spectra were obtained from NGC1068  between 1.25-1.32 microns , in these spectra
have being identified the most important emission lines [FeII]$\lambda$12570, Pa$\beta$ $\lambda$12814 and FeII $\lambda$13201.

For each of the lines, it has made ​​an adjustment of multicomponent obtaining up to five components to the same line in some spectra, which in this work 
only the two most important were presented and mapped: [FeII]$\lambda$12570 , Pa$\beta$ $\lambda$12814 being obtained the velocity field.

The information obtained for the observed distribution of flow in the two mapped components has shown that the emission gas in each has a double origin.
It has being associated with photoionization by young stars, or excitation by collisions. The nuclear emission for [FeII] in both components is greater than Pa$\beta$, 
which means a high influence of shocks produced by the jet; Moreover, the emission intensity of the Pa$\beta$ line is high for the component 1, which imply 
that this component is ionized by a younger cluster, or there are more massive stars. The Map FWHM for [FeII] has higher values ​​for component 1, 
indicating greater gas turbulence in the zone where the line is measured causing deformed wider lines by turbulence. For Pa$\beta$ line, the FWHM obtained
for component 1  is greater than for the component 2. We conclude that for this line, there is greater gas turbulence shown in componte 1 than component 2.

We Obtained for the case of eqw of [FeII]$\lambda$12570 similar data values ​​for the two components show slightly lower in the component 2; For the component two of the
Pa$\beta$ line, values ​​shown slightly lower than for the component 1. Because the component 1 shows a greater Eqw of Pa$\beta$  we can estimate that it would 
occur in an area with more recent star formation or abundant. On the map shown, marked with x, the nuclear region  has values ​​around 2. This would imply
that the shocks, the SN, and the winds dominate there; but moving away from this area, we can not say much due to poor signal to noise and greater errors.
The map for the second component is not shown because  values ​​indicate very low signal to noise ratio and increased errors. 

Finally, from the analysis of the velocity field is obtained that the component one of [FeII] line  is for greater than component 2. 
In the same way, the velocity is higher in component one than component two  for Pa$\beta$ line.
The data relating to the component 1 are consistent with the presence of rotation. However, it is hard to 
distinguish this effect produced by a jet. In the event that the rotation is displaced from the maximum continuous, our observations may point to the presence of a bar.

\section*{Acknowledgements}
This work has been supported by the Universidad Autonoma de Madrid fellowship 2010-2011.
A special  gratitude to  Luc Binette from UNAM for the support in my short stay with him.
My special  gratitude to German Gimeno from Gemini Telescope for the amazing experience at the telescope.
A special appreciation to Antonio Sanchez Ibarra, Carmen Mendiola, Hector P. Coiffier, Rocio P.
Mendiola, \mbox{K. Paola Tellez (and Pity).} 
My gratitude to Luis Felipe Rodriguez from IRYA, Sergio Martin from ESO, Damian Mast from the Observatorio Astronomico de Cordoba, Angeles Diaz from
Universidad Autonoma de Madrid, and Rolf Guesten from Max Planck for Radioastronomy for believing in me.
%%%%%%%%%%%%%%%%%%%%%%%%%%%%%%%%%%%%%%%%%%%%%%%%%%



%\bibliographystyle{mnras}


%%%%%%%%%%%%%%%%%%%%%%%%%%%%%%%%%%%%%%%%%%%%%%%%%%

%%%%%%%%%%%%%%%%% APPENDICES %%%%%%%%%%%%%%%%%%%%%

%\appendix



%%%%%%%%%%%%%%%%%%%%%%%%%%%%%%%%%%%%%%%%%%%%%%%%%%

% Don't change these lines
\bsp	% typesetting comment
\label{lastpage}
\end{document}